\documentclass{article}

\usepackage{arxiv}
\usepackage[utf8]{inputenc} 
\usepackage[T1]{fontenc}    
\usepackage{hyperref} 
\usepackage{url}            
\usepackage{booktabs}       
\usepackage{amsfonts}       
\usepackage{nicefrac}       
\usepackage{microtype}      
\usepackage{lipsum}		
\usepackage{graphicx}
\usepackage{natbib}
\usepackage{doi}
\usepackage{subfigure}

\title{Near Real-time CO$_2$ Emissions Based on Carbon Satellite And Artificial Intelligence}

\author{%
  Zhengwen Zhang$^{1}$, Jinjin Gu$^{2}$, Junhua Zhao$^{1,3,}$\thanks{Corresponding author}~, Jianwei Huang$^{1,3}$, Haifeng Wu$^{4}$\\
  $^{1}$School of Science and Engineering, The Chinese University of Hong Kong, Shenzhen, China\\
  $^{2}$School of Electrical and Information Engineering, The University of Sydney, Sydney, Australia\\
  $^{3}$Shenzhen Institute of Artificial Intelligence and Robotics for Society (AIRS), Shenzhen, China\\
  $^{4}$Shenzhen Finance Institute, Chinese University of Hong Kong, Shenzhen, China
}

\begin{document}
\maketitle

\begin{abstract}
To limit global warming to pre-industrial levels, global governments, industry and academia are taking aggressive efforts to reduce carbon emissions.
The evaluation of anthropogenic carbon dioxide (CO$_2$) emissions, however, depends on the self-reporting information that is not always reliable. 
Society need to develop an objective, independent, and generalized system to meter CO$_2$ emissions.
Satellite CO$_2$ observation from space that reports column-average regional CO$_2$ dry-air mole fractions has gradually indicated its potential to build such a system. 
Nevertheless, estimating anthropogenic CO$_2$ emissions from CO$_2$ observing satellite is bottlenecked by the influence of the highly complicated physical characteristics of atmospheric activities.
Here we provide the first method that combines the advanced artificial intelligence (AI) techniques and the carbon satellite monitor to quantify anthropogenic CO$_2$ emissions.
We propose an integral AI based pipeline that contains both a data retrieval algorithm and a two-step data-driven solution.
First, the data retrieval algorithm can generate effective datasets from multi-modal data including carbon satellite, the information of carbon sources, and several environmental factors.
Second, the two-step data-driven solution that applies the powerful representation of deep learning techniques to learn to quantify anthropogenic CO$_2$ emissions from satellite CO$_2$ observation with other factors.
Our work unmasks the potential of quantifying CO$_2$ emissions based on the combination of deep learning algorithms and the carbon satellite monitor.
\end{abstract}

\section{Introduction}
Human activities contribute to climate change by changing the amount of greenhouse gases, aerosols (fine particulate matter) and clouds in the atmosphere. 
The biggest contributor is the burning of fossil fuels \cite{olivier2005recent}, which emit carbon dioxide (CO$_2$) gas into the atmosphere.
Reducing these emissions is a core objective of the Paris Agreement of the United Nations Framework Convention on Climate Change (UNFCCC).
The Paris Agreement requires all parties to report anthropogenic greenhouse gas emissions and removals at least every 2 years.
In addition, an increasing number of corporates are required to provide annual report for CO$_2$ emission accounts.

Although there are well-established self-reporting mechanisms, we still need alternative carbon emission observation methods to provide validation for the reported data and eliminate potential biases.
Quantifying anthropogenic (CO$_2$) emissions at the individual facility level has important implications.
It can help to monitor emissions reductions or support regulation of carbon trading/pricing systems or other mitigation strategies.
At present, most existing carbon emission estimation methods are based on corporates’ self-reporting information \cite{gurney2021under}, regional or sector level carbon emission factors \cite{shan2016new}, and other publicly available statistics data \cite{shan2020china}. 
However, these methods need extra verification and the estimation frequency is unable to meet the requirements for real-time carbon emission estimation.
In some industrial sectors, continuous emission monitoring systems (CEMS) are being deployed for accurate and real-time carbon emission estimation \cite{jahnke1997handbook}.
But considering their high cost, it is impractical to build a comprehensive carbon emission monitoring system based on CEMS.

Recently, carbon satellite monitoring is a new technical means that can directly provide CO$_2$ column-average dry-air mole fractions (XCO$_2$), which denotes the column-average regional CO$_2$ concentration in atmosphere.
Owing to the potential of supporting CO$_2$ emissions estimation, carbon satellite monitoring has been a fast-growing research topic.
Previous studies qualified CO$_2$ emissions by building a Gaussian plume model to simulate the CO$_2$ flux movement in the atmosphere \cite{nassar2017quantifying}.
In this paper, to estimate carbon emissions from carbon satellite observations at the individual facility level, we propose a pure data-driven methods based on a novel deep learning algorithm.
To the best of our knowledge, this is the first carbon measurement method based on carbon satellite data and state-of-the-art artificial intelligence technologies. 
We anticipate this work to open up new research paradigms in carbon measurement.

Our research is based on the observations of NASA's Orbiting Carbon Observatory 2 (OCO-2) satellite \cite{crisp2017orbit}.
OCO-2 makes high spectral resolution measurements of reflected solar radiation at wavelengths in the 0.76, 1.61, and 2.06 $\mu$m regions to derive XCO$_2$.
OCO-2 has limited imaging capabilities, it measures XCO$_2$ in 8 parallelogram footprints (each is about 1.29 $\times$ 2.25 km$^2$) over a narrow swath (less than 10.3 km).

CO$_2$ plumes derived from large emission sources may cause local enhancement in the OCO-2 observational near-source data because of diffusion and flow of gases, separating it from the background XCO$_2$.
This local enhancement recorded in satellite data is viewed as reflecting the patterns of CO$_2$ emissions and subsequent movement.
Thus we can utilize this pattern to estimate the emissions of carbon source.
Detecting this enhancement and establishing the map from near-source satellite data to CO$_2$ emissions are central to this task.
%
%
We choose to design a deep neural network suitable for carbon satellite's unique data structure based on a Transformer architecture \cite{vaswani2017attention}.
After well-training, given a range of carbon satellite data with local enhancement, carbon source location, and some environmental information, this neural network directly predicts carbon emissions for the location it queried.

However, several challenges make this work not a naive utilization of deep learning on carbon satellite data.
\begin{itemize}
    \item 
    First, the OCO-2 has only limited imaging capabilities, and its data is in the form of discrete measurement locations and the carbon concentrations measured at that location.
    This presents a challenge to the design of deep learning methods.
\end{itemize}
\begin{itemize}
    \item 
    Second, the excess XCO$_2$ generated by large emission sources typically reaches 1\% at the best, which is about 4ppm compared with an instrument noise typically around 0.3--0.6ppm.
    This non-negligible noise in the XCO$_2$ measurement hampers the imaging and detection of emission plumes and the precision of emission quantification.
    In the case of severe measurement noise, it is not even possible to confirm whether there is significant local enhancement with a single measurement.
    Similar to the case of carbon measurements, it is also extremely difficult to obtain an accurate wind field situation, which is another important basis for quantifying the plumes caused by carbon emissions.
    Usually, we can only obtain average wind speed and direction data over a large spatial scale and over a long period of time.
    And this only provides a very limited clue for estimating the emission plume.
\end{itemize}
\begin{itemize}
    \item 
    Third, data acquisition is also challenged due to space satellites' unique measurement orbital limitations.
    Only on rare occasions do the OCO-2 tracks cross CO$_2$ plumes downwind of large cities or power plants, limiting the possibility of quantifying the corresponding CO$_2$ emissions to a few cases within a year.
    This also limits our possibilities for multiple measurements of the same carbon source.
    By matching the location of known carbon sources, wind direction, and the location of the satellite detection swath, we are only able to match the hundreds of available data from millions of OCO-2 records, which also suffer from noise.
\end{itemize}
\begin{itemize}
    \item 
    The last challenge is the complexity of emission source data.
    In this work, we select real hourly emission sources from continuous emission monitoring system (CEMS) data for our research.
    However, not only are the emissions data recorded by CMES facing missing and inaccurate challenges, but the situation of their emission sources is sometimes extremely complex.
    Multiple closely located emission sources may influence each other, making estimation more difficult.
\end{itemize}
To solve above problem, this work makes a three-step data retrieval algorithm and a two-step data-driven solution based on AI.
The data retrieval algorithm is designed to find effective satellite data that contains local enhancement and to  match auxiliary information geographically and temporally.
We apply three steps to achieve this goals:(1) retrieval based on carbon sources for extracting satellite data near carbon source; (2) processing of abnormal data for filtering extremely noisy data; (3) retrieval based on pattern detection for finding effective satellite data with local enhancement.
During data processing, we find that there is scarcity of data with real emission label. 
In contrast, we have massive amount of data without matched real emission label.
%
%
To sufficiently utilize the properties of retrieved multi-modal data, We propose a novel data-driven solution.
Our solution includes two steps: (1) masked pre-training for carbon emission estimation on large-scale of data without real emission label; (2) linear probing for the final prediction on a small-scale data with real emission labels.

The proposed data-driven method makes two important technical advances.
Firstly, we present a new network architecture called CarbonNet for carbon emission estimation from OCO-2 XCO$_2$ measurements.
CarbonNet abandons the traditional neural network design paradigm using fully-connected layers or convolutional layers and adopts a novel strategy based on Transformer.
The OCO-2 measurements can be viewed as independent measurements in different locations.
We treat each measurement as a token containing the location information, the XCO$_2$ number, the global wind information and the statistical background XCO$_2$ number.
In our CarbonNet, these tokens calculate self-attention and interact with each other, i.e. each token can interact with others, greatly improving its computing efficiency.
More importantly, in CarbonNet, the order of tokens is variable, and tokens can be masked flexibly.
This enables our second technical design called mask pre-training.
Recall that of the vast amount of available satellite data, only a small fraction can be matched with recorded emissions sources.
We have very little satellite data with emission labels, but a lot of data without labels.
The distribution of these data is complex and full of noise.
Learning directly with a small amount of data will undoubtedly lead to severe overfitting.
To address this problem, we propose a masked pre-training method using satellite data that cannot match the emission record.
For data without labels, we randomly mask a portion of it, e.g., 25\% of the measurement points, and train the CarbonNet to predict the masked measurements based on the data provided.
This forces the network to learn the distribution of satellite-measured XCO$_2$ and can learn an efficient representation of the satellite data.
Finally, we use linear regression to predict carbon emissions from satellite data representations on the labeled data.
Our experiments show that the proposed solution can effectively predict carbon emissions and can avoid overfitting and noise data interference to a certain extent.

\section{Data}
\label{sec:headings}
In this section, we first introduce three types of raw data and their sources. Then we discuss the provided three-steps data retrieval algorithm in detail. Finally we give the information of generated dataset.
\subsection{Data Acquisition}
We collect three different modal data that contributes to CO$_2$ emissions estimation: (1) carbon satellite observations; (2) the position and emissions of carbon source; (3) some environmental information that influences CO$_2$ movements. 
Firstly, we use version 9r of the OCO-2 bias-corrected XCO$_2$ retrievals, which is a level-2 satellite product that records XCO$_2$ and a lot of supplementary information.
We extract several key recordings from this data, including position, time, XCO$_2$, XCO$_2$ evaluation parameters, and recording angles of the carbon satellite.
The position information is consist of the center and corner coordinates (in the geographic coordinate system) of each scan area. 
XCO$_2$ is the most important key, which reflects the column-averaged regional concentration of carbon dioxide in the atmosphere detected by the carbon satellite.
Two XCO$_2$ evaluation parameters in OCO-2 product (\texttt{xco$_2$\_quality\_flag} and \texttt{xco$_2$\_uncertainty}) are used to evaluate the quality of XCO$_2$ recordings. 
Besides, we choose \texttt{solar\_zenith\_angle} and \texttt{sensor\_zenith\_angle} in recordings that can help us select worth data.
The detailed explanations of these selected keys can be referred to \cite{eldering2017orbiting}. 
We could use these parameters both to find effective data and to estimate the emissions of carbon sources.
On the other hand, we choose real hourly emission data from continuous emissions monitors (CEMS) released by EPA Air Markets Program Data \cite{CEMS}.
It covers over 1303 power plants in the United States and their positions.  
Finally, some environmental information, including the average wind speed of u and v direction below 50m, solar radiation and surface pressure, is derived from the ERA5 reanalysis dataset \cite{C3S}.
\subsection{Data Retrieval}
There are two reasons to develop a data retrieval algorithm before building model to estimate emissions. The first reason is that the quality of data is low, which indicates the low signal-noise rate of satellite observations, incomplete CEMS emissions recordings and low resolution of environmental data. The second reason is that our inversion framework relies on multi-modal data to jointly link the observed CO$_2$ local enhancement with upwind local emission source. Matching of these data significantly limit the the number of effective data.
We did a lot of analysis and filtering of the data. 
The data retrieval is mainly composed of three steps: (1) retrieval based on carbon source; (2) processing of abnormal data; (3) retrieval based on pattern (local enhancement in satellite data) detection.
The details of this three-step data retrieval algorithm are depicted as following.
\subsubsection{Retrieval Based on Carbon Sources}
The first step is to filter some unreliable data by the information of carbon source.
We first merge adjacent power plants to one carbon source if they are too closer to others (the distance between two emission sources is less than 3 km that is the spatial resolution of OCO-2 carbon satellite).
For instance, in New York state, a power plant named Equus Power I is merely 0.7km from another named Freeport Power Plant No. 2. 
Thus, the least resolution of carbon source in our framework is equal to the spatial resolution of carbon satellite. 
After the combination, we obtain 1245 independent carbon sources from 1303 CEMS power plants with position information.
We match all satellite data within $1^{\circ}\times 1^{\circ}$ (longitude $\times$ latitude) wide moving windows centered on each independent carbon source. 
Then these satellite data in each window are divided into 63931 single satellite swath data by different recording times. 
Finally, we remove some satellite swath data if the distance between carbon source and satellite swath is greater than 40 km.
This can increase the probability that the detected local enhancement within satellite swath data is associated with a specific carbon source. 
We have 46179 remains of swath data after retrieval based on carbon sources.
\subsubsection{Processing of Abnormal Data}
Although previous works \cite{nassar2017quantifying, zheng2020observing} choose to use the good-quality data (\texttt{xco$_2$\_quality\_flag} equals 0), we use both the good-quality and bad-quality data (\texttt{xco$_2$\_quality\_flag equals} 1) in our framework. 
Conversely, we encode the quality flag of XCO$_2$ as part of the input of our model. 
The most important reason to do this is to balance the number and quality of the remaining data after data retrieval.
Utilizing all quality data can significantly increase the number of effective data and make subsequent supervised learning possible.
However, it also introduces a lot of extremely noisy recordings in swath data which is harmful for training model. 
Thus, the second step is to filter abnormal satellite swath data according to their statistical information.

We first control the number of effective recording points in each swath data. 
Figure \ref{data_processing}(a) presents the distribution of the length of swath data after retrieval based on carbon source. 
We delete those satellite swath data if their length are less than 200 since they have a tendency to lack key information for emission estimation.
In addition, the maximum length is restricted to 450 to avoid a rare case that a carbon satellite repeatedly scans the same area over a continuous period. 
As shown in Figure \ref{data_processing}(c), we find that there are some outliers in swath data which are extremely large than the mean XCO$_2$. 
These noisy recordings seemly caused by wrong recording or satellite retrieval error in carbon satellite Level 2 production, can significantly decrease the estimation ability of our model. 
Therefore, we directly remove those swath data if the maximum of recording XCO$_2$ is greater than 1000 ppm. 
Finally, we exclude the extremely complex emission situations caused by multiple carbon sources within the matched area.   
Figure \ref{data_processing}(e) shows the distribution of the number of carbon sources within each area.
There may exist almost 40 carbon sources in a $1^{\circ}\times 1^{\circ}$ area (longitude $\times$ latitude), which seems impossible to estimate the emissions of a single carbon source. 
Consequently, we limit the number of carbon sources in each area to 10 to lessen the difficulty of emission estimation. 
Finally, it is necessary to delete extremely slim satellite swaths (the weight of the swath is less than 8 km) generated by inappropriate satellite recording angles.
The corresponding distributions after retrieval based on statistics are shown in Figure \ref{data_processing}(b), \ref{data_processing}(d) and \ref{data_processing}(f) respectively.
After processing of abnormal data, we eventually have about 10930 denoised swath data. 
These swaths can be used to form large-scale data without carbon emission labels for pre-training based on self-supervised learning.

\subsubsection{Retrieval based on Pattern Detection}
The above two steps have helped us to get high-quality swath data. 
However, these swaths may not carry on effective and sufficient information to support emission estimation of carbon sources. 
To avoid learning from data that inputs are irrelevant to outputs, we develop the third step to find out these satellite swath data that contains patterns for emission estimation.
Previous works \cite{nassar2017quantifying, zheng2020observing} tend to fit on a Gaussian plume detected on a swath to estimate corresponding emissions.
However, we merely obtain four swaths if limiting to classical Gaussian plumes on our data.
Therefore, we design a specific pattern detection approach to ensure that the swath data is associated with a single carbon source.

\begin{figure*}[!t]
   \begin{center}
   \includegraphics[width=\linewidth]{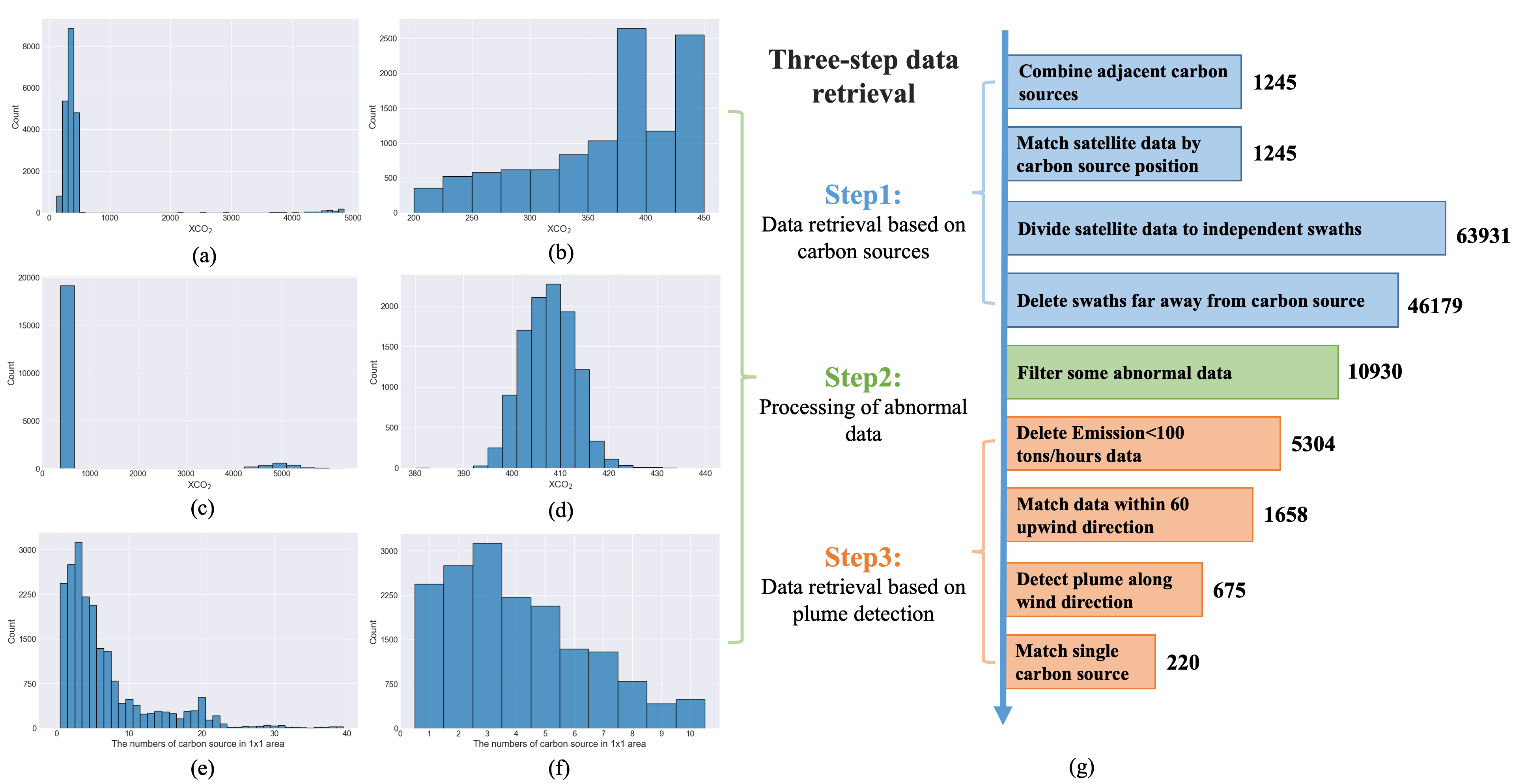}
   \end{center}
      \caption{Before processing of abnormal data (step2): (a) Number of effective points distribution; (c) Max XCO$_2$ distribution;  (e)Number of carbon source in $1^{\circ}\times 1^{\circ}$ (longitude $\times$ latitude) area.After processing of abnormal data: (b) Number of effective points distribution; (d) Maximum XCO$_2$ distribution;  (f)Number of carbon source in $1^{\circ}\times 1^{\circ}$ area. (g) The remaining cases after three-step retrieval based on carbon source, statistics and pattern detection respectively.}
   \label{data_processing}
   \end{figure*}

We first filter these swath data either without recorded CEMS emissions or with matched emissions less than 100 tons/hours, and get 5304 remaining swath data.
We set the maximum upwind direction to 60 as in \cite{zheng2020observing} to ensure that the CO$_2$ movement from a carbon source is reflected in carbon satellite swath data in term of local enhancement. 
Merely 1658 swath data are successfully matched within 60 upwind direction.
We fit a curve to XCO$_2$ retrieval data along the orbit which is centered on the intersection of swath data and carbon source along upwind direction.
These swaths are considered as carrying potential local enhancemnet if the values of their intersection along wind direction exceed the standard deviation of spatial variability above the local average within 200km.
Additionally, we remove some cases where several carbon sources contribute to the same potential XCO$_2$ local enhancement. 
Only 220 effective swath data have been left after all filters.
The number of cases remained after each filtering shown in Figure \ref{data_processing}(g).
\subsection{Datasets}
According to above three-step data retrieval method, we have a large-scale unlabeled dataset and a small-scale real-emission labeled data.
The unlabeled dataset that contains 10930 swath data is collected for pre-training to extract effective and efficient features. 
The light labeled swath data include only 220 swath data, we random choose 150 of them as training data used for fitting a linear model, and the other 70 of them are for testing performance. 
\section{Method}
In this section, we describe our method in detail. First, we introduce the input data encoding and deep architecture of the proposed novel transformer-based CarbonNet. After that we provide a two-step solution to learn to estimate CO$_2$ emissions. The first step adopt the mask pre-training strategy to learn sufficient representation on the large-scale unlabeled satellite data. We extract the deep features from well mask pre-training model on a small-scale real-emission labeled data for downstream prediction. The second step take the combination of deep features and some handcraft features on a small-scale real-emission labeled data as input, and learns a weighted linear model for predicting CO$_2$ emissions. 

\subsection{The Transformer-based CarbonNet}
We propose a novel deep neural network architecture namely CarbonNet and associated data encoding approach. 
We first describe the process of encoding satellite swath data to the form for training. 
The XCO$_2$ measurements from the OCO-2 satellite can be viewed as a collection of tuples $\{(lon_i, lat_i, x_i)\}_{i=1}^{n_j}$, where $lon_i$ and $lat_i$ represents the longitude and the latitude of the $i$th measurement point, $x_i$ is the XCO$_2$ record at this point and $n_j$ is the number of measurements in swath data $j$.
As described in Section 2, $n_j$ is a number between 200 and 450 in our paper. 
We incorporate the wind information and other statistic information of swath into each tuple to provide the supporting information for prediction.
The average wind below 50m for each point is denoted as $(u, v)$, where $u$ and $v$ is the value of wind speed of the longitude and latitude components.
We also incorporate some statistic information that includes the means and standard deviation XCO$_2$ of each swath, and the background XCO$_2$. 
The statistic information record is denoted as $(m_i, s_i, b_i)$.
For different data at the same time, the background carbon dioxide concentration $b_i$ is assumed to be a constant as mean $m_i$ and standard deviation $s_i$.
Because carbon dioxide concentrations vary primarily with time, they have increased periodically over the past few years.

\begin{figure*}[!t]
   \begin{center}
   \includegraphics[width=\linewidth]{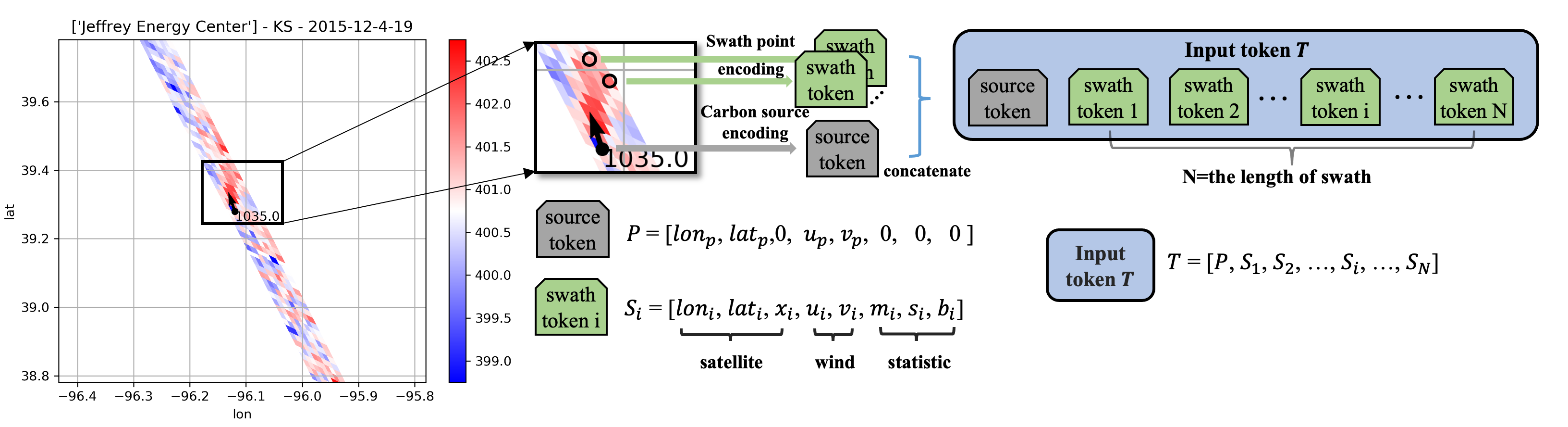}
   \end{center}
      \caption{The process of encoding a swath data to the form used for training.}
   \label{encoding}
   \end{figure*}

In the proposed CarbonNet, we treat each tuple as a special vector namely token.
We reorganize the data in each tuple as a vector $S_i=[lon_i, lat_i, x_i, u_i, v_i, m_i, s_i, b_i]$, named swath token, and the whole swath data is represented as a sequence of vectors $S'\in\mathbb{R}^{n_j\times8}$.
For subsequent training CarbonNet, each vector sequence $S'$ is padded with 0 to the same length N that is set to the maximum length of swath data (450 in our paper).  
Therefore, each swath is encoded as a sequence of vectors with constant length, denoted as $S\in\mathbb{R}^{N\times8}$.
We also encode the position and wind information of carbon source as a vector $P=[lon_p, lat_p, 0, u_p, v_p, 0, 0, 0]$, where $lon_p$ and $lat_p$ denote the longitude and latitude of carbon source, $u_p$ and $v_p$ denote the u and v wind at the position of carbon source.
$P$ is named source token as the same length as each swath token. 
Finally, we use the the concatenation of source token and a sequence of swath token $T=[P,S]\in\mathbb{R}^{(N+1)\times8}$ as input token of CarbonNet.
A case of the input data encoding is as shown in Figure \ref{encoding}.

We start to introduce the architecture of proposed CarbonNet.
Given a swath input $T$ we use a fully-connected embedding layer $E(\cdot)$ to extract feature $F_0\in\mathbb{R}^{n\times c}$ as
$$F_0=E(T),$$
where $c$ is the dimension of the feature.
Then, we extract deep feature $F_{DF}\in\mathbb{R}^{n\times c}$ as
$$F_{DF}=H_{DF}(F_0),$$
where $H_{DF}(\cdot)$ is the deep feature extraction module and it
contains K residual building blocks.
More specifically, intermediate features $F_1$, $F_2$,$\dots$, $F_K$ and the output deep feature $F_{DF}\in\mathbb{R}^{c}$ are extracted block by block as
$$F_i=H_i(F_{i-1}),\quad F_{DF}[i]=\frac{1}{n}\sum_{j=1}^n F_K[ji],$$
where $H_i(\cdot)$ denotes the $i$th building block and $F_{DF}$ is obtained by taking the average over the $n$ dimension.

Each building block has a self-attention layer $H_{SA}$ in \cite{vaswani2017attention} and a fully-connected layer $H_{FC}$.
In the self-attention layer of the $i$th building block, the query $Q$, key $K$ and value $V$ are computed as
$$Q= F_{i-1}W^{Q}, K=F_{i-1}W^{K}, V=F_{i-1}W^{V},$$
where $W^{Q},W^{K},W^{V}\in\mathbb{R}^{c\times d}$ are weight matrices, and $d$ is the number of projected vectors.
Then, we use $Q$ to query $K$ to generate the attention map $A\in\mathbb{R}^{n\times n}$ as
$$A=\mathtt{softmax}( \frac{QK^T}{\sqrt{D}}+ B),$$
where $B$ is the learnable relative positional encoding.
This attention map $A$ is then used for the weighted sum of $n$ vectors in $F_{i-1}$.

Next, a fully-connected layer with GELU non-linearity activation is used for further feature transformations as
$$H_{FC}(F;W^{FC})=FW^{FC}+b^{FC},$$
where $W^{FC}$ is the weight and $b^{FC}$ is the bias.
The LayerNorm (LN) layer is added before both $H_{SA}$ and $H_{FC}$, and the residual connection is employed for both modules.
The whole process is formulated as
$$F_{intermediate} = H_{SA}(LN(F_{i-1})) + F_{i-1}$$
$$F_{i}=H_{FC}(LN(F_{intermediate})) + F_{intermediate}.$$
\subsection{Masked Pre-training for Carbon Emission Estimation}
We next describe our method for masked pre-training strategy.
Recall that we only have very limited labeled data, and a large number of unlabeled data.
The masked pre-training method is to learn a meaningful representation for the downstream prediction through a mask-prediction pretext task.
We assume a binary 0-1 mask $M\in\mathbb{R}^{N\times c}$ that randomly masks each token with a probability $\rho$, i.e. the values of masked tokens will be replaced by 0.
This ratio $\rho$ can be large, depending on the redundancy of the data.
For images, we can even safely mask out up to 90\% of the pixels.
For our OCO-2 satellite data, we can mask out 25\% -- 50\% of the data to get the best pre-training results as shown in \ref{mask_pretrain}.
The masked swath data is represented as $T_M=T\odot M$ and is sent into the CarbonNet to get the $K$th intermediate features $F_K$.
We use another projector layer $Proj(\cdot)$ to project the $c$ dimensional features back to the 9-dimensional tokens as
$$T'=Proj(F_K)=F_KW^{Proj}+b^{Proj},$$
where the $W^{Proj}$ is the projection weight, and $b^{Proj}$ is the projection bias.
During the pre-training, we input the masked swath data $T_M$ into the CarbonNet and get the projected (predicted) tokens $T'=Proj\circ H_K\circ H_{K-1}\circ \dots H_1\circ E(T_M)$.
We optimize the $l_1$ distance between $T'$ and the unmasked input $T$ to complete the swath data based on part of known information.
In this way, even in the absence of labeled data, CarbonNet still learns the patterns in the swath data and extracts a high-quality representation $F_K$.
After pre-training, the projection layer $P$ is discarded, but the rest of the parameterized part is reserved for feature extraction.

\subsection{Linear Probing for the Final Prediction}
\begin{figure}[!t]
\centering
\begin{minipage}{0.4\linewidth}
\centering
\subfigure[]{
\label{mask_pretrain}
\centering
\includegraphics[width=\linewidth]{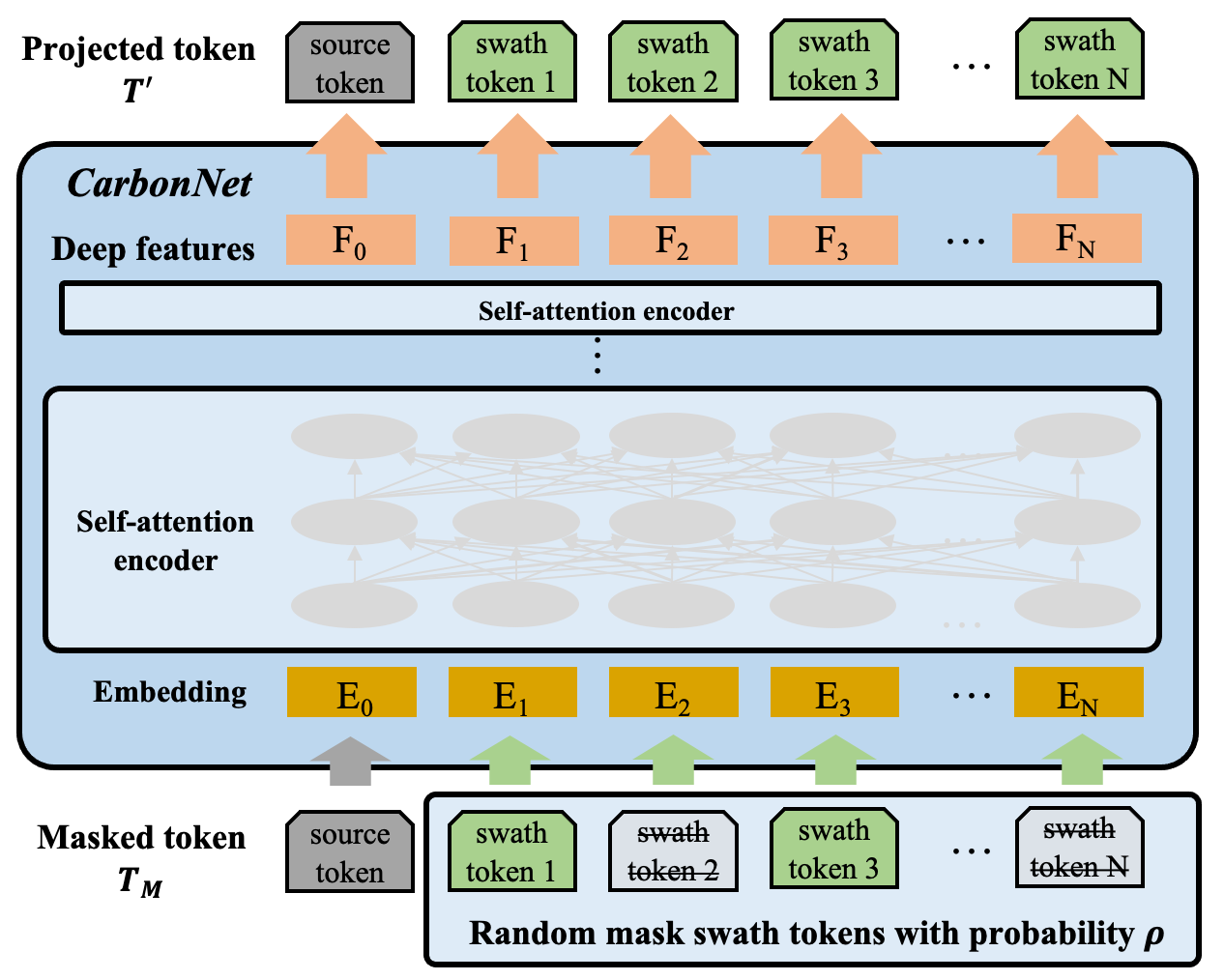}}
\end{minipage}\qquad
\begin{minipage}{0.5\linewidth}
\centering
\subfigure[]{
\label{linear_regression}
\centering
\includegraphics[width=\linewidth]{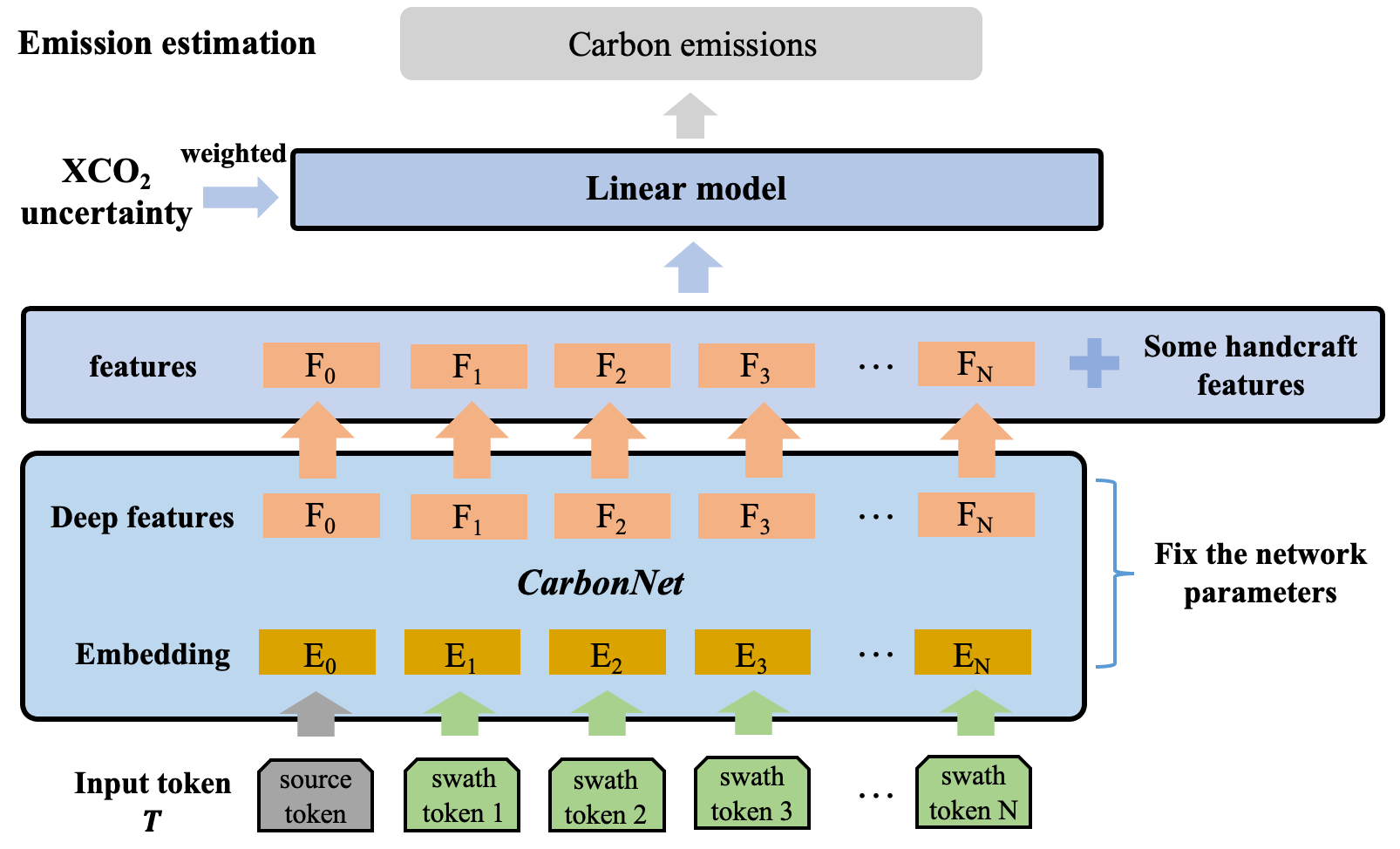}}
\end{minipage}
\caption{(a) The framework of pre-training on large scale of unlabeled data. (b) Linear regression on small scale of labeled data by deep features extracted from CarbonNet.}
\label{method}
\end{figure}

The above pre-training strategy gives us a well-trained deep representation of OCO-2 swath data, we can simply obtain the carbon estimation results using the linear probing method as shown in Figure \ref{linear_regression}.
Linear Probing refers to fixing the features of the representation layer and training the classifier or regressor only through supervised data.
Given the well-trained CarbonNet, we extract the deep features $F_{DF,k}\in\mathbb{R}^c, k\in\{1,2,\dots,N\}$ for each labeled data in our dataset.
Then we stack extracted deep features with some handcraft features, including the mean and deviation of swath, the position and wind information of carbon source, and the background XCO$_2$ of swath.
The final input features is a $c+d$ dimensional vector, where $d$ denotes the dimension of handcraft features.
Combined with the carbon emissions records matched by these data, we obtain the paired data to learn a linear regression weighted by XCO$_2$ uncertainty between the c-dimensional feature and the carbon emissions as
$$e_k=F_{DF,k}W^{e} + b^e,$$
where $W^{e}\in\mathbb{R}^{c+d}$ is the linear weight, and $b^e$ is the bias term.
The weights can be obtained in a variety of ways, in this work we simply use linear regression to train.

\section{Results}
To estimate CO$_2$ emissions from single carbon source, we use multi-modal data, including some parameters from version 9r of the OCO-2 bias-corrected XCO$_2$ retrievals, hourly real CO$_2$ emissions and power plant information from CEMS data, and wind data collected from ERA5-land. 
We provide three-step data retrieval method to generate effective satellite data that contains patterns. 
Then we propose a two-step solution to estimate CO$_2$ emissions.
The first step can extract sufficient deep features of satellite data from the layer before the final dense layer of the well-trained model. 
The trend of training loss and the visualization of recovery satellite data can demonstrate the effectiveness of mask pre-training.
Second, we learn a weighted linear model, which takes the combination of deep features and handcraft features as input and directly outputs the estimated emissions of given carbon source. 
Our experiments show that the proposed method can effectively predict CO$_2$ emissions and be robust to noisy data to a certain extent. 
Finally, we compared the proposed solution with an existing method called Gaussian Plume Model (GPM). 
The conducted experiments show that our method can outperform GPM on the test set.
\subsection{Representation Extraction of CarbonNet Based on Masked Pre-training}
\begin{figure*}[!h]
\begin{center}
\includegraphics[scale=0.3]{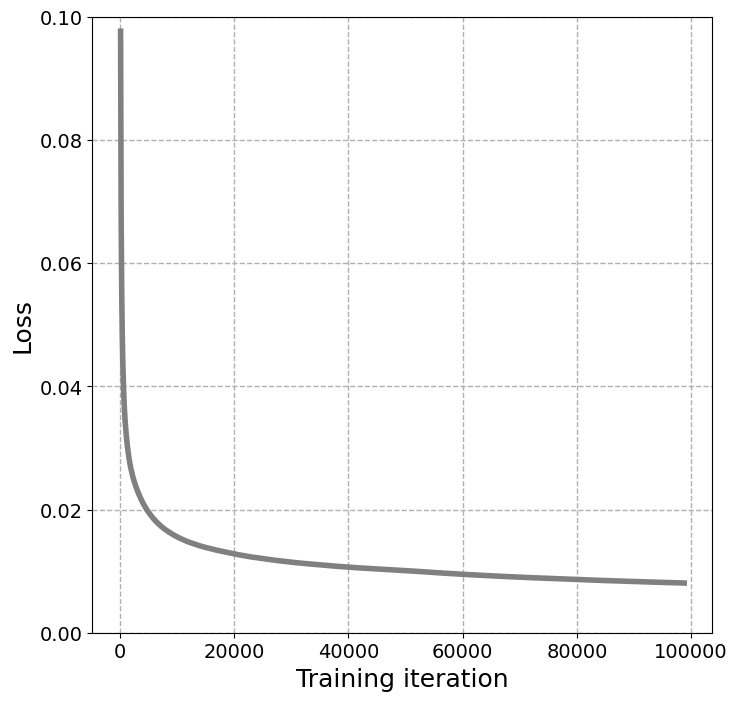}
\end{center}
\caption{The trade of L1 loss descending with training iterations.}
\label{l1_loss}
\end{figure*}

\begin{figure*}[!t]
\centering
\begin{minipage}{0.3\linewidth}
\centering
\subfigure[]{
\label{original_swath}
\centering
\includegraphics[width=0.9\linewidth]{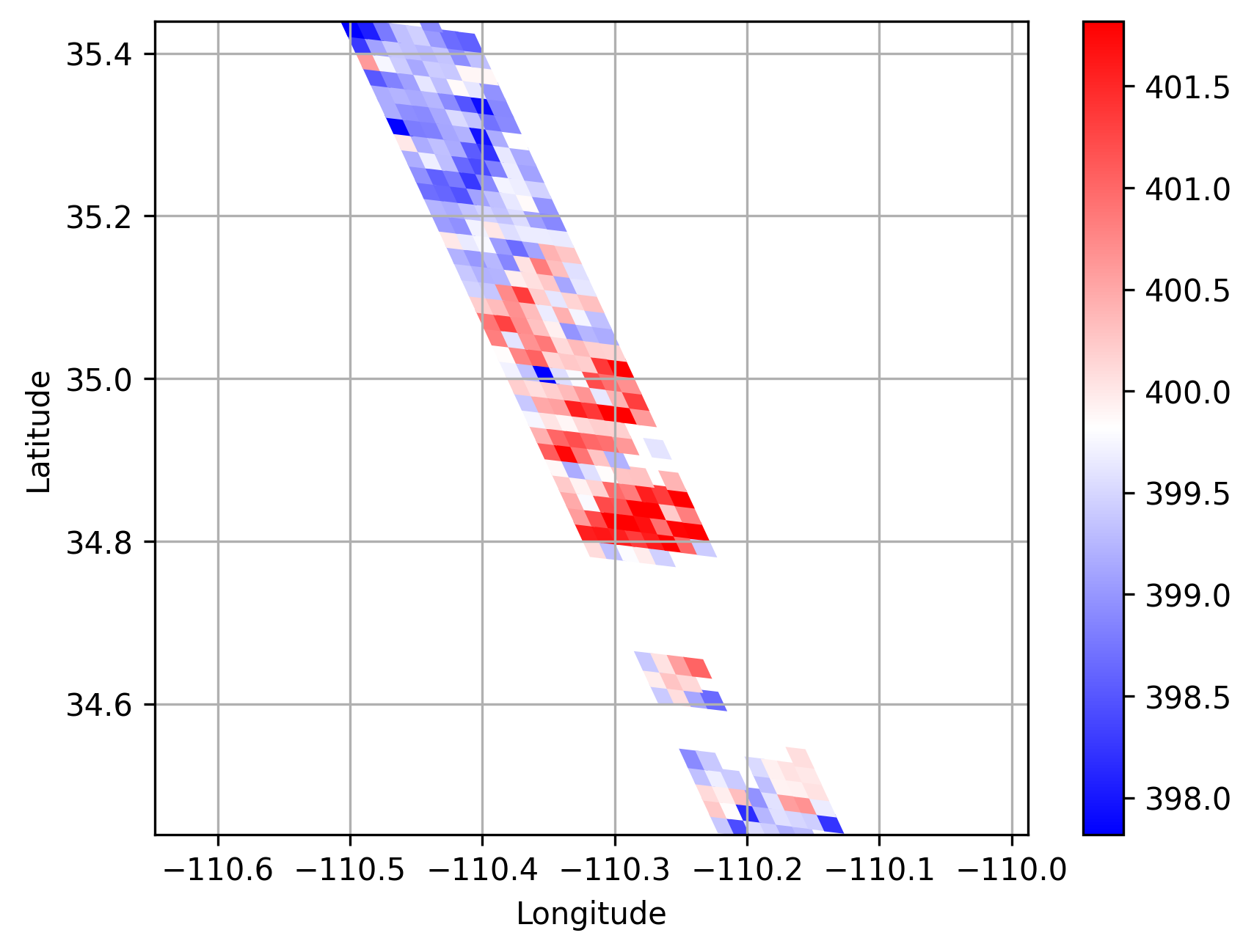}}
\end{minipage}
\begin{minipage}{0.3\linewidth}
\centering
\subfigure[]{
\label{masked_swath}
\centering
\includegraphics[width=0.9\linewidth]{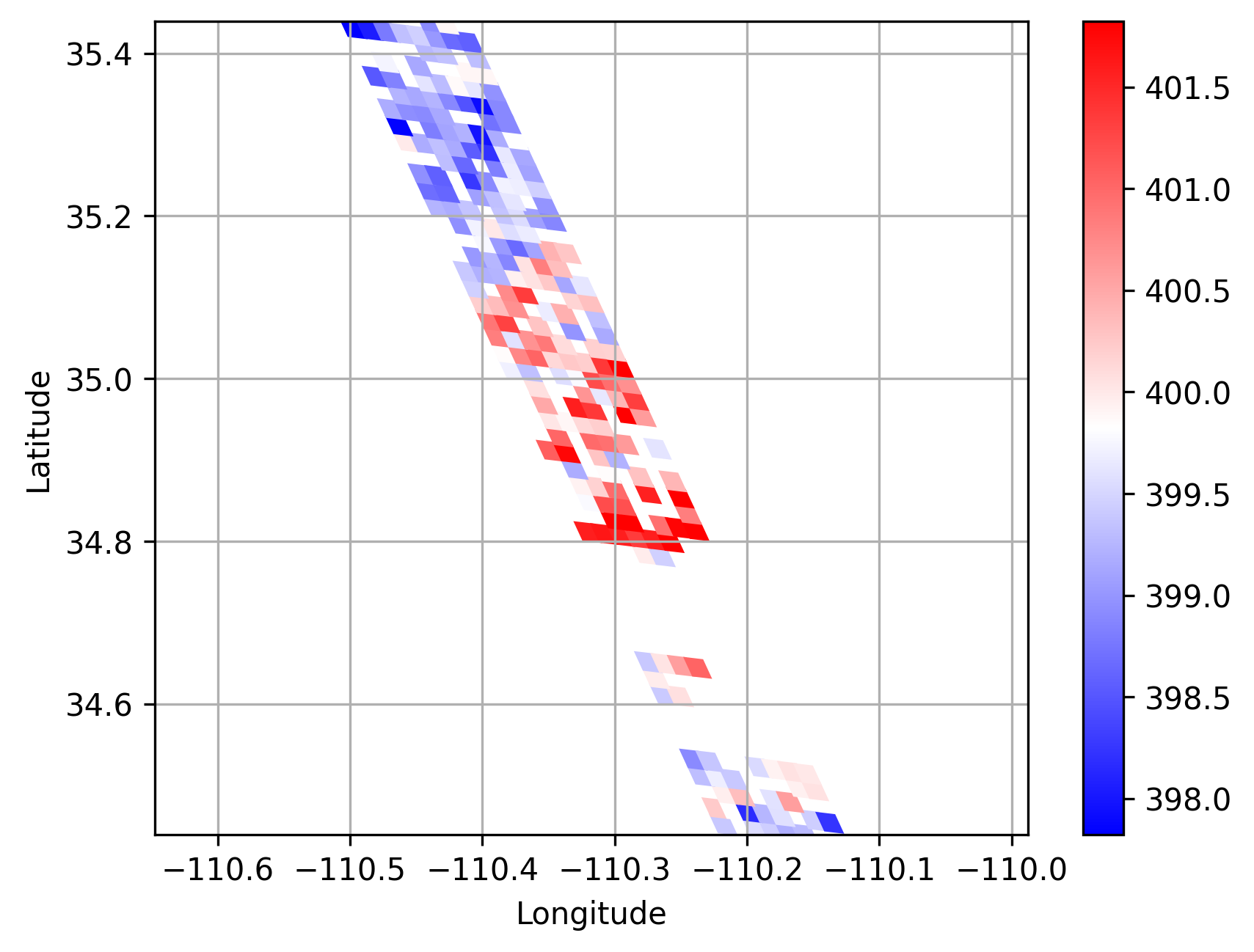}}
\end{minipage}
\begin{minipage}{0.3\linewidth}
\centering
\subfigure[]{
\label{recovered_swath}
\centering
\includegraphics[width=0.9\linewidth]{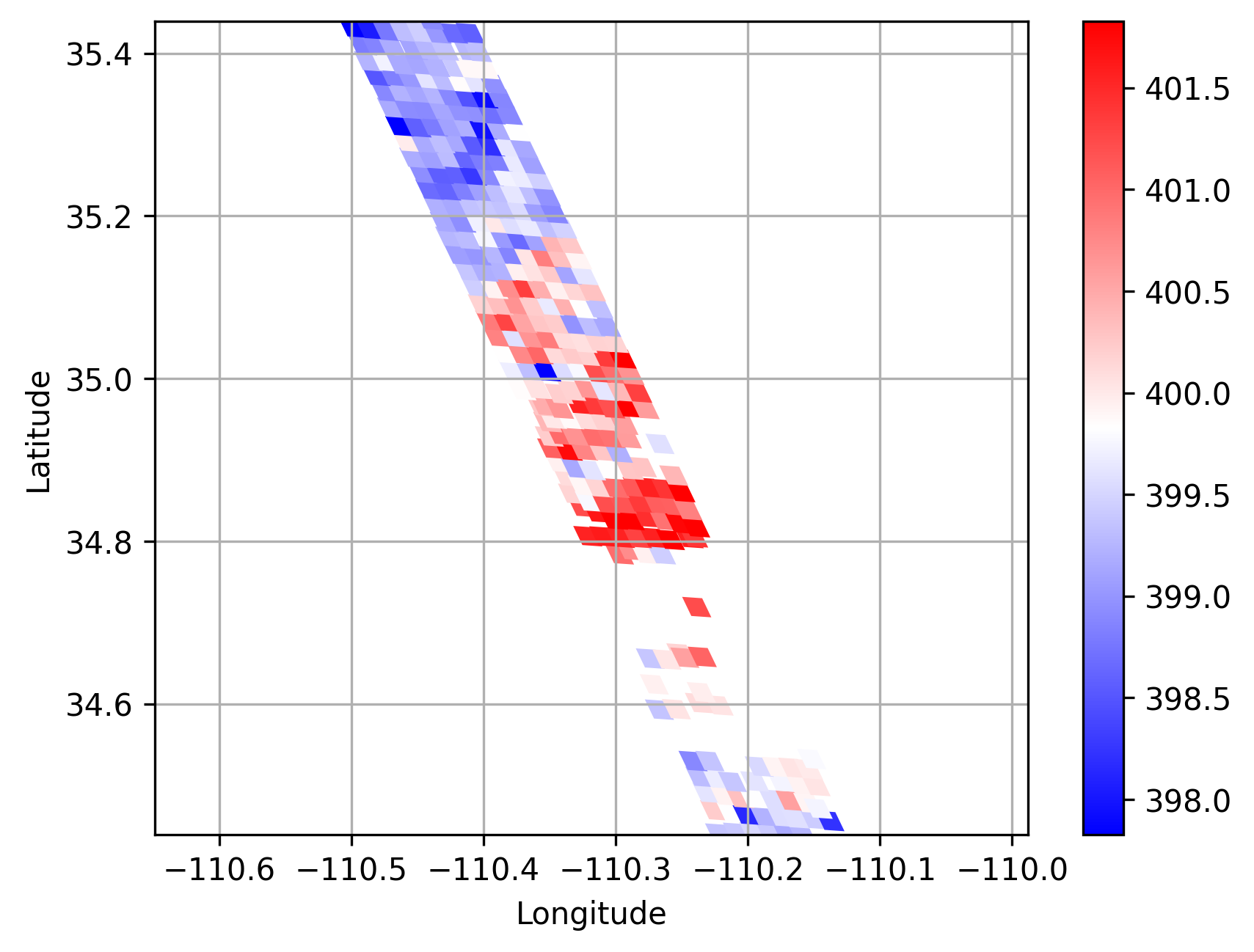}}
\end{minipage}
\caption{(a) Originally satellite data. (b) 25\% Randomly masked satellite data. (c) Retrieved data from masked input.}
\label{visual_comp}
\end{figure*}

In the mask pre-training stage, we set the probability of random mask to 25\%. 
Figure \ref{l1_loss} shows the L1 loss descending trend with the training iterations.
It is easy to see that our approach, after 100000 iterations, achieves the lowest Loss value equals about 0.8\% over all 5304 unlabeled satellite data.
This demonstrates that our pre-trained method has the ability to recover masked swath information from the unmasked parts.
Thus, the features extracted from pre-trained CarbonNet can serve as an efficient representation of the satellite swath data.
Figure \ref{visual_comp} tries to provide a case of visualized comparison between original swath data and the recovered data predicted by pre-trained CarbonNet.
On 25 February 2015 at about 20 local time in state Arizona near a power plant named Cholla, the original satellite XCO$_2$ data, masked data and retrieved data from masked input are presented in Figure \ref{original_swath}, \ref{masked_swath} and \ref{recovered_swath} respectively.
Intuitively, we obtain the same conclusion that the masked parts are successfully recovered in the retrieved satellite data.
\subsection{CO$_2$ emissions estimates by Linear Regression}
\begin{figure}[!h]
\centering
\begin{minipage}{0.32\linewidth}
\centering
\subfigure[]{
\label{net_results}
\centering
\includegraphics[width=\linewidth]{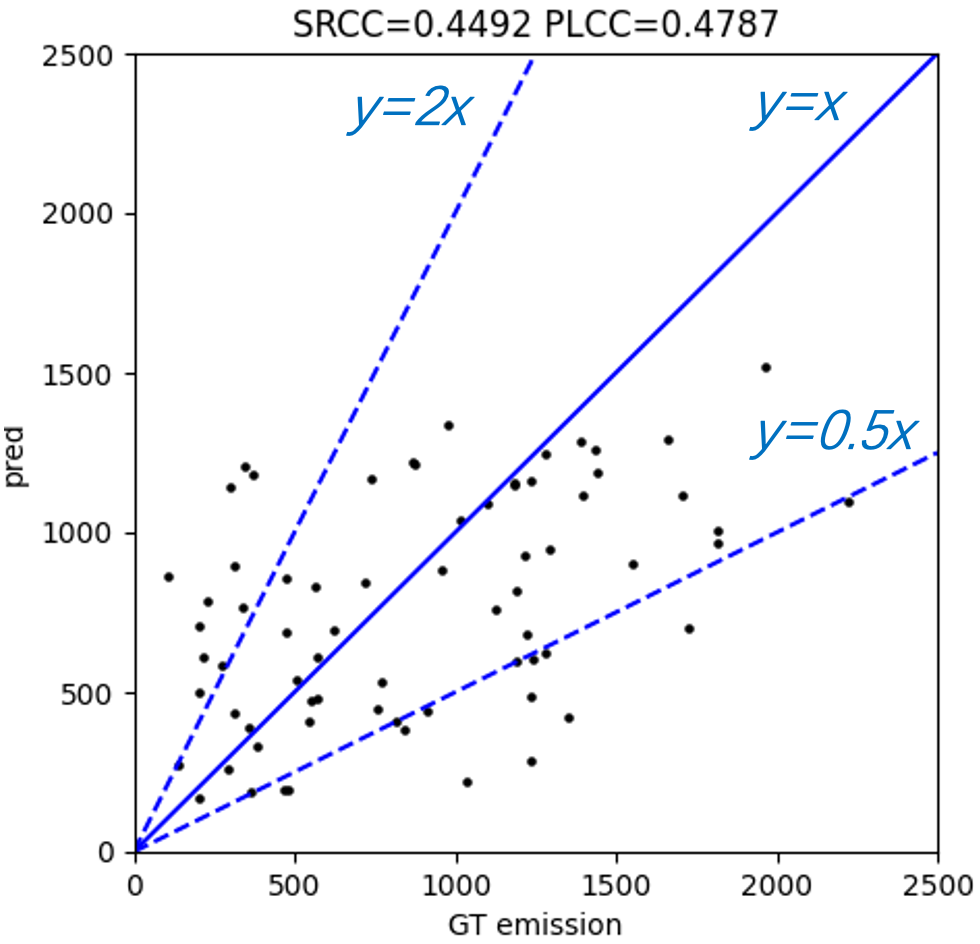}}
\end{minipage}\qquad
\begin{minipage}{0.32\linewidth}
\centering
\subfigure[]{
\label{GPM_results}
\centering
\includegraphics[width=\linewidth]{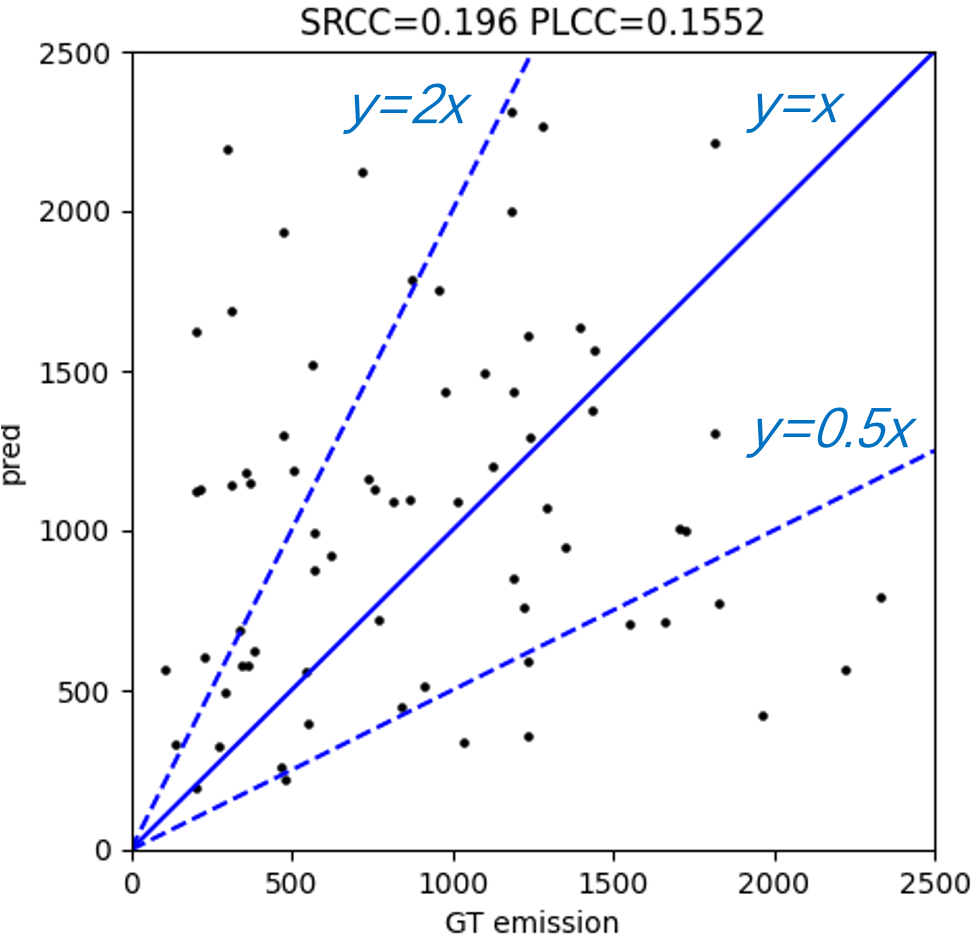}}
\end{minipage}
\caption{(a) The performance of our method on 70 test set. (b) The performance of Gaussian plume model method on 70 test set. The blue line denotes expected $y=x$ and the dotted lines denotes $y=2x$ and $y=0.5x$ respectively.}
\label{results}
\end{figure}

\begin{figure}[!h]
\centering
\begin{minipage}{0.4\linewidth}
\centering
\subfigure[]{
\label{vis_inclassical_GP}
\centering
\includegraphics[width=\linewidth]{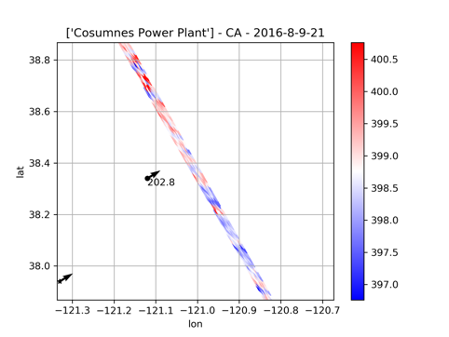}}
\end{minipage}\qquad
\begin{minipage}{0.4\linewidth}
\centering
\subfigure[]{
\label{inclassical_GP}
\centering
\includegraphics[width=\linewidth]{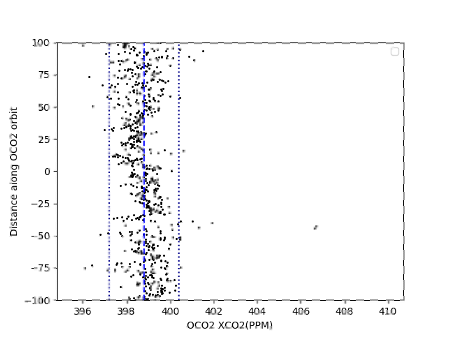}}
\end{minipage}
\caption{(a) The visualization of a case of atypical Gaussian plume. (b) Distance along OCO$_2$ orbit to XCO$_2$ (PPM) of a case of atypical Gaussian plume.}
\label{case_results}
\end{figure}

In contrast to pre-training on large-scale swath data, a weighted linear model is learned on small-scale labeled swath data to estimate CO$_2$ emissions.
We first extract and save features of each labeled swath data from the layer before the final dense layer of pre-trained CarbonNet.
Then we concatenate these features to some handcraft features that include longitude and latitude of carbon source, u and v direction wind speed at the position of carbon source, the mean and standard deviation of this satellite swath data, monthly background mean and standard deviation.
we fit a linear model weighted by XCO$_2$ uncertainty and XCO$_2$ quality flags which takes concatenated features as input and labeled emissions as output.
Figure \ref{net_results} shows the final test performance of weighted linear regression on 70 test set. 
On the other hand, we test the performance by applying the Gaussian plume model method proposed in \cite{nassar2017quantifying} as shown in Figure \ref{GPM_results}. 
We use Spearman rank-order correlation coefficient (SRCC) and Pearson linear correlation coefficient (PLCC) to evaluate the correlation between the estimation and the ground truth.
The results show that our method can achieve higher correlation coefficients (both SRCC and PLCC) than the Gaussian plume model method. 
Moreover, it is obvious that our method has fewer outliers which are out of the blue lines ($y=2x$ and $y=0.5x$). 
%
As shown in Figure \ref{inclassical_GP}, a case with 202.8 tons/hour emission label that does not an contain explicit Gaussian plume is predicted to be 262.7 tons/hour by our model. 
However, the prediction of fitting Gaussian plume model method is 1184.2 tons/hour which surpasses five-fold true emissions. 

\section{Related Work and Discussion}
Based on the traditional global greenhouse gas ground-based observation network, it is hard to understand the anthropogenic sources and sinks of greenhouse gas. 
Monitoring greenhouse gas from space by carbon satellite has trend to be a critical part of the new generation of carbon metering system.
It is a great challenge to extract anthropogenic emissions data from XCO$_2$ data measured by carbon satellites.
The measurement of gas such as nitrogen dioxide (NO$_2$) and sulfur dioxide (SO$_2$) based on satellite imaging spectrometers has already been demonstrated in \cite{liu2016no, liu2017no}.
However, the precision of CO$_2$ measurement is obviously interior to NO$_2$ or SO$_2$, which brings to a great challenge to quantifying CO$_2$ emissions.
Some studies use synthetically generated CO$_2$ emissions of power plants \cite{bovensmann2010remote} and cities \cite{pillai2016tracking} to prove the ability to perceive CO$_2$ emissions from carbon satellites. 
An alternative is to estimate CO$_2$ emissions by first getting accurate NO$_2$ emissions and then finding the relationship between CO$_2$ emissions and NO$_2$ emissions of a power plant.
Moreover, there are some works to quantify CO$_2$ emissions by fitting real measured emissions from CEMS installed on power plants \cite{nassar2017quantifying}.
Nevertheless, all previous CO$_2$ emission estimation methods are based on establishing an atmospheric physics model.
To our knowledge, no attempt has been made yet to infer CO$_2$ emissions based on the data-driven method before this paper.
As shown above, we propose a two-step data-driven solution to quantifying CO$_2$ emissions from the carbon source.
The main limitation of our method is lack of explicit interpretability as all black-box models.
It aggravates the difficulty in evaluating the error of emission estimation by our model. 
\section{Conclusion}
In this paper, we first reveal the potential of the data-driven AI method to quantify CO$_2$ emissions from space.
We first design a three-step data retrieval algorithm to find effective multi-modal data with patterns that enable to map from satellite data to emissions. 
The provided data retrieval algorithm can generate a large-scale unlabeled data and light-scale data with real emission label matched from CEMS data.
According to the properties of data, we propose a novel deep architecture namely CarbonNet and a two-step solution to utilize both large-scale unlabeled data and small-scale labeled data for CO$_2$ emissions estimation.
The first step based on mask pre-training strategy can obtain sufficient deep features for downstream prediction.
The second step predicts emissions by learning a weighted linear model that takes the combination of deep features and some handcraft features as input.
Owing to the design of encoding multi-modal data as special vector namely token, our method can easily benefit from more accumulated swath data and higher precision data from the new generation of carbon monitors.
For future work, we intend to collect more relevant information, such as the topography of the selected area, temperature, and moisture, to further improve the accuracy of emission estimation of our solutions.

\bibliographystyle{unsrtnat}
\bibliography{references} 

\begin{thebibliography}{16}
\providecommand{\natexlab}[1]{#1}
\providecommand{\url}[1]{\texttt{#1}}
\expandafter\ifx\csname urlstyle\endcsname\relax
  \providecommand{\doi}[1]{doi: #1}\else
  \providecommand{\doi}{doi: \begingroup \urlstyle{rm}\Url}\fi

\bibitem[Olivier et~al.(2005)Olivier, Van~Aardenne, Dentener, Pagliari,
  Ganzeveld, and Peters]{olivier2005recent}
Jos~GJ Olivier, John~A Van~Aardenne, Frank~J Dentener, Valerio Pagliari,
  Laurens~N Ganzeveld, and Jeroen~AHW Peters.
\newblock Recent trends in global greenhouse gas emissions: regional trends
  1970--2000 and spatial distributionof key sources in 2000.
\newblock \emph{Environmental Sciences}, 2\penalty0 (2-3):\penalty0 81--99,
  2005.

\bibitem[Gurney et~al.(2021)Gurney, Liang, Roest, Song, Mueller, and
  Lauvaux]{gurney2021under}
Kevin~Robert Gurney, Jianming Liang, Geoffrey Roest, Yang Song, Kimberly
  Mueller, and Thomas Lauvaux.
\newblock Under-reporting of greenhouse gas emissions in us cities.
\newblock \emph{Nature communications}, 12\penalty0 (1):\penalty0 1--7, 2021.

\bibitem[Shan et~al.(2016)Shan, Liu, Liu, Xu, Shao, Wang, and
  Guan]{shan2016new}
Yuli Shan, Jianghua Liu, Zhu Liu, Xinwanghao Xu, Shuai Shao, Peng Wang, and
  Dabo Guan.
\newblock New provincial co$_2$ emission inventories in china based on apparent
  energy consumption data and updated emission factors.
\newblock \emph{Applied Energy}, 184:\penalty0 742--750, 2016.

\bibitem[Shan et~al.(2020)Shan, Huang, Guan, and Hubacek]{shan2020china}
Yuli Shan, Qi~Huang, Dabo Guan, and Klaus Hubacek.
\newblock China co2 emission accounts 2016--2017.
\newblock \emph{Scientific data}, 7\penalty0 (1):\penalty0 1--9, 2020.

\bibitem[Jahnke(1997)]{jahnke1997handbook}
James~A Jahnke.
\newblock \emph{Handbook, Continuous Emission Monitoring Systems for
  Non-criteria Pollutants}.
\newblock Center for Environmental Research Information, National Risk
  Management~…, 1997.

\bibitem[Nassar et~al.(2017)Nassar, Hill, McLinden, Wunch, Jones, and
  Crisp]{nassar2017quantifying}
Ray Nassar, Timothy~G Hill, Chris~A McLinden, Debra Wunch, Dylan~BA Jones, and
  David Crisp.
\newblock Quantifying co2 emissions from individual power plants from space.
\newblock \emph{Geophysical Research Letters}, 44\penalty0 (19):\penalty0
  10--045, 2017.

\bibitem[Crisp et~al.(2017)Crisp, Pollock, Rosenberg, Chapsky, Lee, Oyafuso,
  Frankenberg, O'Dell, Bruegge, Doran, et~al.]{crisp2017orbit}
David Crisp, Harold~R Pollock, Robert Rosenberg, Lars Chapsky, Richard~AM Lee,
  Fabiano~A Oyafuso, Christian Frankenberg, Christopher~W O'Dell, Carol~J
  Bruegge, Gary~B Doran, et~al.
\newblock The on-orbit performance of the orbiting carbon observatory-2 (oco-2)
  instrument and its radiometrically calibrated products.
\newblock \emph{Atmospheric Measurement Techniques}, 10\penalty0 (1):\penalty0
  59--81, 2017.

\bibitem[Vaswani et~al.(2017)Vaswani, Shazeer, Parmar, Uszkoreit, Jones, Gomez,
  Kaiser, and Polosukhin]{vaswani2017attention}
Ashish Vaswani, Noam Shazeer, Niki Parmar, Jakob Uszkoreit, Llion Jones,
  Aidan~N Gomez, {\L}ukasz Kaiser, and Illia Polosukhin.
\newblock Attention is all you need.
\newblock \emph{Advances in neural information processing systems}, 30, 2017.

\bibitem[Eldering et~al.(2017)Eldering, Wennberg, Crisp, Schimel, Gunson,
  Chatterjee, Liu, Schwandner, Sun, O’dell, et~al.]{eldering2017orbiting}
A~Eldering, PO~Wennberg, D~Crisp, DS~Schimel, MR~Gunson, A~Chatterjee, J~Liu,
  FM~Schwandner, Y~Sun, CW~O’dell, et~al.
\newblock The orbiting carbon observatory-2 early science investigations of
  regional carbon dioxide fluxes.
\newblock \emph{Science}, 358\penalty0 (6360):\penalty0 eaam5745, 2017.

\bibitem[(EPA)()]{CEMS}
United States Environmental Protection~Agency (EPA).
\newblock Washington, dc:office of atmospheric programs, clean air markets
  division.
\newblock \url{https://campd.epa.gov/}.
\newblock 27 June 2022.

\bibitem[(C3S)()]{C3S}
Copernicus Climate Change~Service (C3S).
\newblock Era5: Fifth generation of ecmwf atmospheric reanalyses of the global
  climate, copernicus climate change service climate data store (cds).
\newblock \url{https://cds.climate.copernicus.eu/cdsapp#!/home}.
\newblock last access: 8 September 2022.

\bibitem[Zheng et~al.(2020)Zheng, Chevallier, Ciais, Broquet, Wang, Lian, and
  Zhao]{zheng2020observing}
Bo~Zheng, Fr{\'e}d{\'e}ric Chevallier, Philippe Ciais, Gr{\'e}goire Broquet,
  Yilong Wang, Jinghui Lian, and Yuanhong Zhao.
\newblock Observing carbon dioxide emissions over china's cities and industrial
  areas with the orbiting carbon observatory-2.
\newblock \emph{Atmospheric Chemistry and Physics}, 20\penalty0 (14):\penalty0
  8501--8510, 2020.

\bibitem[Liu et~al.(2016)Liu, Beirle, Zhang, D{\"o}rner, He, and
  Wagner]{liu2016no}
Fei Liu, Steffen Beirle, Qiang Zhang, Steffen D{\"o}rner, Kebin He, and Thomas
  Wagner.
\newblock No$_x$ lifetimes and emissions of cities and power plants in polluted
  background estimated by satellite observations.
\newblock \emph{Atmospheric Chemistry and Physics}, 16\penalty0 (8):\penalty0
  5283--5298, 2016.

\bibitem[Liu et~al.(2017)Liu, Beirle, Zhang, Van Der~A, Zheng, Tong, and
  He]{liu2017no}
Fei Liu, Steffen Beirle, Qiang Zhang, Ronald~J Van Der~A, Bo~Zheng, Dan Tong,
  and Kebin He.
\newblock No$_x$ emission trends over chinese cities estimated from omi
  observations during 2005 to 2015.
\newblock \emph{Atmospheric Chemistry and Physics}, 17\penalty0 (15):\penalty0
  9261--9275, 2017.

\bibitem[Bovensmann et~al.(2010)Bovensmann, Buchwitz, Burrows, Reuter, Krings,
  Gerilowski, Schneising, Heymann, Tretner, and Erzinger]{bovensmann2010remote}
H~Bovensmann, M~Buchwitz, JP~Burrows, M~Reuter, T~Krings, K~Gerilowski,
  O~Schneising, J~Heymann, A~Tretner, and Joerg Erzinger.
\newblock A remote sensing technique for global monitoring of power plant
  co$_2$ emissions from space and related applications.
\newblock \emph{Atmospheric Measurement Techniques}, 3\penalty0 (4):\penalty0
  781--811, 2010.

\bibitem[Pillai et~al.(2016)Pillai, Buchwitz, Gerbig, Koch, Reuter, Bovensmann,
  Marshall, and Burrows]{pillai2016tracking}
Dhanyalekshmi Pillai, Michael Buchwitz, Christoph Gerbig, Thomas Koch,
  Maximilian Reuter, Heinrich Bovensmann, Julia Marshall, and John~P Burrows.
\newblock Tracking city co$_2$ emissions from space using a high-resolution
  inverse modelling approach: a case study for berlin, germany.
\newblock \emph{Atmospheric Chemistry and Physics}, 16\penalty0 (15):\penalty0
  9591--9610, 2016.

\end{thebibliography}
\end{document}